\documentclass[twocolumn,superscriptaddress,aps,prb,longbibliography]{revtex4-2}
\usepackage{graphicx}% Include figure files
\usepackage{dcolumn}% Align table columns on decimal point
\usepackage{bm}% bold math
\usepackage{amssymb, amsmath}
\usepackage{xcolor}
\usepackage{multirow}
\usepackage{slashed}

\begin{document}

\title{Reflection-enhanced gain in traveling-wave parametric amplifiers}

	\author{S.~Kern}
	\affiliation{Department of Experimental Physics, Comenius University, SK-84248 Bratislava, Slovakia}
	
	\author{P.~Neilinger}
	\affiliation{Department of Experimental Physics, Comenius University, SK-84248 Bratislava, Slovakia}
	\affiliation{Institute of Physics, Slovak Academy of Sciences, D\'{u}bravsk\'{a} cesta, Bratislava, Slovakia}

	\author{E.~Il'ichev}
	\affiliation{Leibniz Institute of Photonic Technology, D-07702 Jena, Germany}
	
	\author{A.~Sultanov}
	\affiliation{Department of Applied Physics, Aalto University, P.O. Box 15100, FI-00076 Aalto, Finland}

	\author{M.~Schmelz}
	\affiliation{Leibniz Institute of Photonic Technology, D-07702 Jena, Germany}
	
    \author{S.~Linzen}
	\affiliation{Leibniz Institute of Photonic Technology, D-07702 Jena, Germany}

	\author{J.~Kunert}
	\affiliation{Leibniz Institute of Photonic Technology, D-07702 Jena, Germany}
	
	\author{G.~Oelsner}
	\affiliation{Leibniz Institute of Photonic Technology, D-07702 Jena, Germany}
	
	\author{R.~Stolz}
	\affiliation{Leibniz Institute of Photonic Technology, D-07702 Jena, Germany}
	
	\author{A.~Danilov}
	\affiliation{Department of Microtechnology and Nanoscience MC2, Chalmers University of Technology, SE-41296 Goteborg, Sweden}

	\author{S.~Mahashabde}
	\affiliation{Department of Microtechnology and Nanoscience MC2, Chalmers University of Technology, SE-41296 Goteborg, Sweden}

	\author{A.~Jayaraman}
	\affiliation{Department of Microtechnology and Nanoscience MC2, Chalmers University of Technology, SE-41296 Goteborg, Sweden}

	\author{V.~Antonov}
	\affiliation{Physics Department, Royal Holloway, University of London, Egham TW20 0EX, United Kingdom}
	
	\author{S.~Kubatkin}
	\affiliation{Department of Microtechnology and Nanoscience MC2, Chalmers University of Technology, SE-41296 Goteborg, Sweden}
	
	\author{M.~Grajcar}
	\affiliation{Department of Experimental Physics, Comenius University, SK-84248 Bratislava, Slovakia}
	\affiliation{Institute of Physics, Slovak Academy of Sciences, 	D\'{u}bravsk\'{a} cesta, Bratislava, Slovakia}

\begin{abstract}
The operating principle of traveling-wave parametric amplifiers is typically understood in terms of the  standard
coupled mode theory, which describes the evolution of forward propagating waves without any
reflections, i.e. for perfect impedance matching. However, in practice, superconducting microwave
amplifiers are unmatched nonlinear finite-length devices, where the reflecting waves undergo complex
parametric processes, not described by the standard coupled mode theory. Here, we present an analytical solution for the TWPA gain, which includes the interaction of
reflected waves. These reflections result in corrections to the well-known results of the standard coupled mode theory, which are obtained for both 3-wave and 4-wave mixing processes. Due to these reflections,
gain is enhanced and unwanted nonlinear phase modulations are suppressed. Predictions of
the model are experimentally demonstrated on two types of unmatched TWPA, based on coplanar waveguides
with a central wire consisting of i) a high kinetic inductance superconductor, and ii) an array of 2000
Josephson junctions.
\end{abstract}
\pacs{}
\maketitle

%%%%%%%%%%%%%%%%%%%%
\section{Introduction}
\label{sec:intro}
%%%%%%%%%%%%%%%%%%%%

Modern microwave quantum engineering exploits efficient detection of low power microwave signals
\cite{krantz2019quantum,ili2004} requiring linear amplification with ultra-low added noise.
Nowadays, superconducting parametric amplifiers, exhibiting quantum limited sensitivity, have
become the most favourable implementation in practical devices \cite{devor}.

The superconducting parametric amplifiers that have been demonstrated so far can be divided in two classes: the
resonant parametric amplifier \cite{yurke1987squeezed,yurke1989observation,yurke1996low,
zagoskin2008generation} and the traveling wave parametric amplifier (TWPA)
\cite{eom2012wideband,white2015traveling,macklin2015near,planat2020photonic}. A resonant parametric amplifier
works as a nonlinear resonator, which provides energy transfer from a strong pump tone to the signal to be
amplified \cite{tholen2007nonlinearities}. The finite interaction time of the waves providing amplification is
enhanced by a high quality factor of the nonlinear resonator, which in turn limits the bandwidth.
Typically, the required nonlinearity is achieved by integrating the resonator with an appropriate
array of Josephson junctions (JJs) \cite{castellanos2007widely}.

In a TWPA, the 3-wave mixing (3WM) (in the presence of DC bias) or the 4-wave mixing (4WM) process occurs, as
waves propagate in a relatively long nonlinear transmission line (TL) \cite{vissers2016low}. The performance
of the TWPA is usually described by coupled mode equations (CME) in the standard form for three types of waves:
pump, signal, and idler, analogously to the fiber optics theory \cite{nonlinear_fiber}. Standard CME
predict that the bandwidth is limited by nonlinear phase modulations only. However, these modulations can be
controlled by means of dispersion engineering. For instance, in Ref. \onlinecite{white2015traveling}, smoothed
broadband amplification from 4 to 8 GHz has been achieved via 4WM and in \cite{malnou2021three} from 3.5 to 5.5
GHz under 3WM.

Analysis based on conventional CME, however, does not take into account counter-propagating waves reflected at the impedance mismatch between the nonlinear TL and standard $50~\Omega$ circuitry. Despite the decade-long lasting effort, the development of an impedance matched device is still a challenge. In experiments, an impedance mismatch results
in modulation of the amplifier transmission (ripples) \cite{chaudhuri2017broadband,goldstein2020four}.  This modulation can be significant, even more than $5~$dB \cite{eom2012wideband,white2015traveling},
which limits the applicability of these amplifiers.
So far, these ripples have been described as Fabry-P\'{e}rot-like resonances
\cite{planat2020photonic,zhao2021quantum} with bandwidth inversely proportional to the length
of the waveguide. Despite having significant influence on the amplifier's transmission, the role of these resonances in parametric processes, to our best
knowledge, has not been described satisfactorily.

Modifications of CME, required for photonic crystal engineering, account for reflections inside the photonic
medium composing the TWPA \cite{huang1994coupled,christopoulos2018degenerate,erickson2017theory}.
In this paper, we generalize the conventional CME for both 3WM and 4WM processes,
by taking into account the reflections at the ends of unmatched TWPA. This enables us to properly describe
the transmission of broadband TWPAs, including resonance effects, such as gain enhancement analogous to the high quality resonant amplifiers.
We have fabricated two coplanar waveguides, one made of high kinetic inductance (KI) superconductor, one consisting of
2000 JJs, both connected to $50~\Omega$ input and output lines. They were tested in 3WM and 4WM regime,
respectively. A reasonable agreement between theory and experiment is demonstrated. Due to the impedance mismatch, both amplifiers operate in the intermediate regime between
the resonant and the traveling-wave limits.

%%%%%%%%%%%%%%%%%%%%
\section{Coupled mode theory}
%%%%%%%%%%%%%%%%%%%%
Non-linear media are commonly exploited for parametric amplification.
In our case,  such medium is provided by the middle wire of the TL, which is formed
either by an array of JJs (see (b) in Fig.\ref{fig:schemes}) or by a high kinetic inductance
superconductor. Dependencies of voltage $V(z,t)$ and current $I(z,t)$ on the coordinate ($z$) and
time ($t$) in the TL are described by nonlinear telegrapher's equations:
\begin{equation}
\label{eq:TEV}
\frac{\partial V(z,t)}{\partial t}=-\frac{1}{C_l}\frac{\partial I(z,t)}{\partial z},
\end{equation}
\begin{equation}
\label{eq:TEI1}
\frac{\partial I(z,t)}{\partial t}=-(L_l(1+I(z,t)^2/I_\ast^2))^{-1}\frac{\partial V(z,t)}{\partial z},
\end{equation}
where $L_l$ and $C_l$ denote the respective inductance and capacitance per unit length of the TL.
Here $I_\ast$ is scale of nonlinearity and relates to the critical current $I_c$, as will be discussed below. As the superconducting medium does not exhibit DC losses, the
DC bias current $I_D$ does not contribute to the voltage $V$, however, it alters the nonlinear inductance.
Therefore, in the presence of $I_D$, equation (\ref{eq:TEI1}) becomes
\begin{equation}
\label{eq:TEI2}
\frac{\partial I(z,t)}{\partial t}=-(L_l^D(1+\epsilon I(z,t)+\xi I(z,t)^2))^{-1}\frac{\partial V(z,t)}{\partial z},\\
\end{equation}
where $\epsilon  =2I_{D}/(I_{D}^2+I_\ast^2)$, $\xi=1/(I_{D}^2+I_\ast^2)$ \cite{malnou2021three}
and the inductance per unit length enhanced by the DC bias is $L_l^D=L_l(1+I_D^2/I_\ast^2)$. In the following,
$I(z,t)$ denotes only the RF current and $I_D$ is fixed. Equivalently, equations (\ref{eq:TEV}) and
(\ref{eq:TEI2}) can be presented as:
\begin{equation}
\label{eq:WE}
v^2\frac{\partial^2 I(z,t)}{\partial z^2} - \frac{\partial^2 I(z,t)}{\partial t^2} = \frac{\partial^2}{\partial t^2}\Big(\frac{1}{2}\epsilon I(z,t)^2+\frac{1}{3}\xi I(z,t)^3\Big),
\end{equation}
where $v=1/\sqrt{L_l^DC_l}$ is the phase velocity in the waveguide.

To extract the gain as a function of circuit parameters, typically four planar waves
\begin{equation}
\label{eq:waves1}
I(z,t) =\sum_{\substack{n \in \\ \{p,s,i_3,i_4\}}} \frac{1}{2}\Big(\mathcal{I}_n(z)e^{i(k_nz-\omega_n t)} + c.c.\Big)
\end{equation}
are substituted  into the equation (\ref{eq:WE}).
Here $\omega_n$ is the circular frequency and $k_n=\omega_n/v$ denotes the wave vector, where the index $n$ indicates the type of the wave;
namely $n = p, s, i_3,i_4$ stands for pump, signal, and two idlers, respectively. The idler $i_3$ entering the 3WM satisfies
\begin{equation}
\label{eq:EC3}
\omega_p=\omega_s+\omega_{i_3}
\end{equation}
and 4WM idler $i_4$ obeys
\begin{equation}
\label{eq:EC4}
2\omega_p=\omega_s+\omega_{i_4}.
\end{equation}
Commonly, it is assumed that such waves propagate along an ideally matched TL.
In practice, however, impedance mismatches at interconnections of different parts of the TL are present.
This leads to partial reflection of the propagating waves, characterized by the reflection coefficient $\Gamma_n$
determined by the impedance mismatch at the frequency $\omega_n$. Taking such reflections at both ends of the non-linear TL into
account (see Appendix \ref{ap:CM}), the RF current in the TL can be expressed as
\begin{equation}
\label{eq:waves2}
I(z,t) = \sum_{\substack{n \in \\ \{p,s,i_3,i_4\}}}\frac{1}{2}\Big(\mathcal{I}_n(z)t_n(e^{ik_nz} + \Gamma_n e^{-ik_nz})e^{-i\omega_n t} + c.c.\Big),
\end{equation}
where $t_n=1/\left(1-\Gamma_n^2 e^{2ik_nl}\right)$ is the transmission amplitude at the frequency $\omega_n$ and $l$ is
the length of the TL. Therefore, the transmission coefficient can be expressed
\begin{equation}
\label{eq:Tran}
T_n\equiv  (1-\Gamma_n^2)^2\lvert t_n\rvert^2=\frac{(1-\Gamma_n^2)^2}{1+\Gamma_n^4 - 2\Gamma_n^2\textrm{cos}(2k_nl)}.
\end{equation}
To reconstruct the functions $\mathcal{I}_s(z)$, $\mathcal{I}_{i_3}(z)$ and $\mathcal{I}_{i_4}(z)$, the expression (\ref{eq:waves2})
is substituted into the wave equation (\ref{eq:WE}). Here, we emphasize that $\mathcal{I}_n(z)$ in the ansatz (8) can contain rapidly oscillating terms and therefore, the amplitudes of the  resulting forward and backward propagating modes of $I(z,t)$ could have a different spatial shape. However, utilizing the averaging method, we extract only a slowly varying part of the $\mathcal{I}_n(z)$ which mainly contributes to the forward propagation (see. Appendices \ref{ap:CM} and \ref{ch:app}). From now on, we denote by symbol $\mathcal{I}_n(z)$ the slowly varying part of the aforementioned amplitudes only. Differential equations for the spatial evolution of these slowly varying amplitudes describing signal gain under 3WM and 4WM are obtained in an approximate form in the limit $\Gamma\ll1$ (see Appendix \ref{ch:app}). Finally,
a solution formally identical to the well-known equation for signal gain \cite{nonlinear_fiber}
is obtained from \eqref{eq:fin}:
\begin{equation}
\label{eq:Gl}
\mathcal{G}_{j}(l)\equiv\left|\frac{\mathcal{I}_s(l)}{\mathcal{I}_s(0)}\right|^2=\textrm{cosh}^2(g_{j}l)+\frac{\beta_{j}^2}{4g_{j}^2}\textrm{sinh}^2(g_{j}l),
\end{equation}
where $\beta_{j}$ is the parameter of the phase mismatch and $g_{j}$ is the gain factor and index $j=3,4$ indicates 3WM or 4WM, respectively.
It is important to note that expression (\ref{eq:Gl}) is generalized to include reflections,
which result in the following correction to the phase mismatch $\beta_{j}$ and the gain factor $g_{j}$.
The phase mismatch for 3WM process $\beta_3$ takes the form
\begin{equation}
\label{eq:beta3}
\beta_3=\Delta k_3(1+2\gamma(1+\Gamma_p^2))-k_p\gamma(1-\Gamma_p^2),
\end{equation}
where $\Delta k_3 =k_p-k_s-k_{i_3}$ is the mismatch of wave vectors and
\begin{equation}
\label{eq:gamma}
\gamma = \frac{\left|t_p\mathcal{I}_p\right|^2}{8(I_D^2+I_\ast^2)}
\end{equation}
is the strength of the nonlinearity. Now, the phase matching condition, i.e. $\beta_3=0$, dictates:
\begin{equation}
\label{eq:pmc3}
\Delta k_3 =k_p \frac{\gamma(1-\Gamma_p^2)}{1+2\gamma(1+\Gamma_p^2)}.
\end{equation}
Obviously, the presence of reflections suppresses the phase mismatch caused by the nonlinearity.
Moreover, $g_3$ is increased, compared to a perfectly impedance-matched system
with $\Gamma_p=0$:
\begin{equation}
\label{eq:g3}
g_3=\sqrt{k_sk_{i_3}\gamma\frac{8I_D^2}{(I_D^2+I_\ast^2)}(1+\Gamma_p^2)-\frac{\beta_3^2}{4}}.
\end{equation}

The 4WM is obtained for zero DC bias, thus all waves participating in the mixing are partially reflected at the ends of the TWPA.
This leads to even more radical changes. The phase mismatch $\beta_4$ is
\begin{equation}
\label{eq:beta4}
\beta_4=\Delta k_4(1+2\gamma(1+\Gamma_p^2))-2k_p\gamma(1-\Gamma_p^2),
\end{equation}
where $\Delta k_4 =2k_p-k_s-k_i$ is the mismatch of wave vectors under 4WM.
The enhancement of the gain factor for 4WM is
\begin{equation}
\label{eq:g4}
g_4=\sqrt{k_sk_{i_4}\gamma^2(1+4\Gamma_p^2)-\frac{\beta_4^2}{4}}.
\end{equation}

\begin{figure} [h]
	\includegraphics[width=0.9\linewidth]{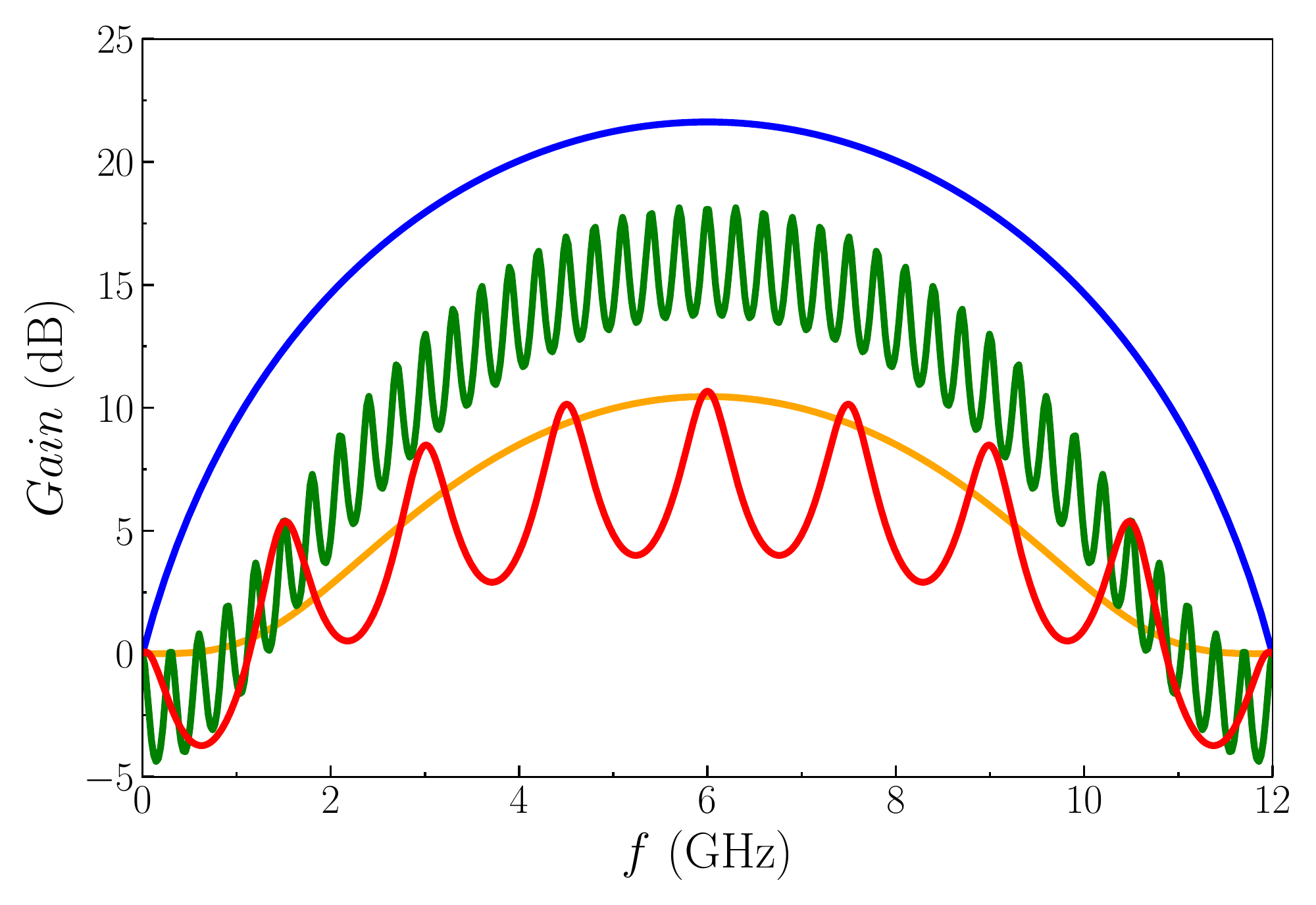}
	\caption{Calculation of 4WM gain using standard CME ($\Gamma_p=0$) for no dispersion engineering (orange curve) and for perfect phase-matching (blue curve). Parameters are taken from \cite{planat2020photonic}. Green curve is prediction of modified theory (Eq. (\ref{eq:Ga})) with $\Gamma_p=0.45$ without any dispersion engineering. Red curve is the modified theory for 5 times shorter TL, no dispersion engineering and higher reflection coefficient $\Gamma_p=0.6$. The gain of the short TL is comparable to standard theory prediction of long TL without phase-matching and the bandwith of each peak is $\approx0.5~$GHz.}
	\label{fig:BvG}
\end{figure}

Both current amplitudes - the amplified signal and the pump entering
the  $\gamma$ coefficient (Eq.~\ref{eq:gamma}) - are modulated by the Fabry-P\'{e}rot-like transmission amplitude $t_n$.
Therefore, the gain of the unmatched amplifier reads
\begin{equation}
\label{eq:Ga}
G_{j}(l)\equiv\left|\frac{I_s(l)}{I_s(0)}\right|^2=\mathcal{G}_{j}(l)T_s, \quad j=3,4.
\end{equation}
To demonstrate the effect of reflections, we took the length and phase velocity from Ref. \onlinecite{planat2020photonic} and plotted the gain curves according to Eq. (\ref{eq:Ga})  in Fig. \ref{fig:BvG} for a demonstrative set of parameters. The resulting curves for the standard CME (or for $\Gamma_p=0$) with and without phase matching are blue and orange, respectively. The green curve depicts the gain, according to Eq. (\ref{eq:Ga}), with $\Gamma=0.45$ without phase matching. Sharp ripples appear and the gain is much higher in comparison to the orange curve. A 5 times shorter waveguide would result in much wider ripples and its gain (red curve) is comparable to the standard theory at original length.

According to the equations (\ref{eq:Gl}-\ref{eq:Ga}) the gain of a TWPA increases with the reflections occurring at the ends of the TL. Vice versa, as is shown in Appendix \ref{app:Q_inc},
the presence of amplification changes the current and voltage wave propagation in such a way, that the reflections at the end of the TL are increased, too. This positive feedback may result in
very high gain in the resonant peaks, whose quality is enhanced (ripples are sharpened). These so-called gain ripples are observed by many groups dealing with TWPA, even for nearly impedance-matched
devices (aiming for $\Gamma_n \to 0$) \cite{goldstein2020four,planat2020photonic}. The effect is not further studied here. In the presented experiment, this effect is observed as well and for measurements with a strong pump, a new value of $\Gamma^\prime$ is introduced.

%%%%%%%%%%%%%%%%%%%%
\section{Experiment}
%%%%%%%%%%%%%%%%%%%%
\subsection{Kinetic inductance TWPA}

%\textbf{Will be completed by Chalmers!}
\begin{figure} [h]
	\includegraphics[width=0.9\linewidth]{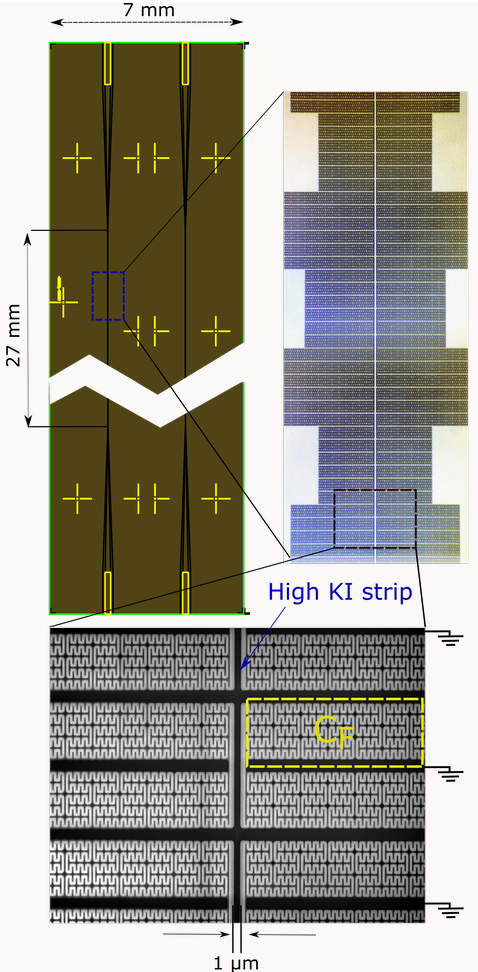}
	\caption{Schematic of the KI-TWPA with chip dimensions. The right inset shows an optical micrograph of a few sections
	of quasi-fractal waveguide. The bottom inset is a magnified SEM image of the fractal structure. The width of the central high
	KI line is 1 $\mu$m. }
	\label{fig:KI}
\end{figure}

In order to investigate the validity of the model described above, we fabricated the KI-TWPA device shown in
Fig~\ref{fig:KI}. The design represents a waveguide where an inductive element - the central high-KI strip - is coupled to
the ground plane via fractalized capacitors C$_F$  (Fig~\ref{fig:KI}- bottom inset). The central strip is 27 mm long and 1
$\mu$m wide and has a sheet kinetic inductance of $\sim$ 4.2~pH/$\square$.
The combination of a highly inductive central strip and increased capacitance of fractalized capacitors results in slow propagation
velocity ($\sim$2\% of the speed of light), thereby reducing the length of the transmission line required to get sufficient amplification \cite{graaf2012magnetic, adamyan2016superconducting}. 	
Moreover, the width of the capacitors is periodically varied along the line to achieve dispersion engineering, with stop band at $\approx10$~GHz. In the vicinity of the bandgap, the phase propagation velocity is perturbed, so that with a proper choice of the pump frequency the dispersion phase shift $\Delta k_3$ compensates for nonlinear phase shift acquired by high power pump and the $\beta=0$ condition can be fulfilled \cite{eom2012wideband}.

The KI-TWPA chip was measured in a Helium gas-flow cryostat with a base temperature of $\sim$ 2 K. The input and
output terminals of the device were bonded onto sampling lines in a printed circuit board (PCB) and the microwave
transmission was measured using a Vector Network Analyzer (VNA). In order to eliminate spurious ground plane
resonances, the ground plane of the chip was carefully bonded to the PCB ground around the perimeter of the sample.

\begin{figure}
	\includegraphics[width=0.99\linewidth]{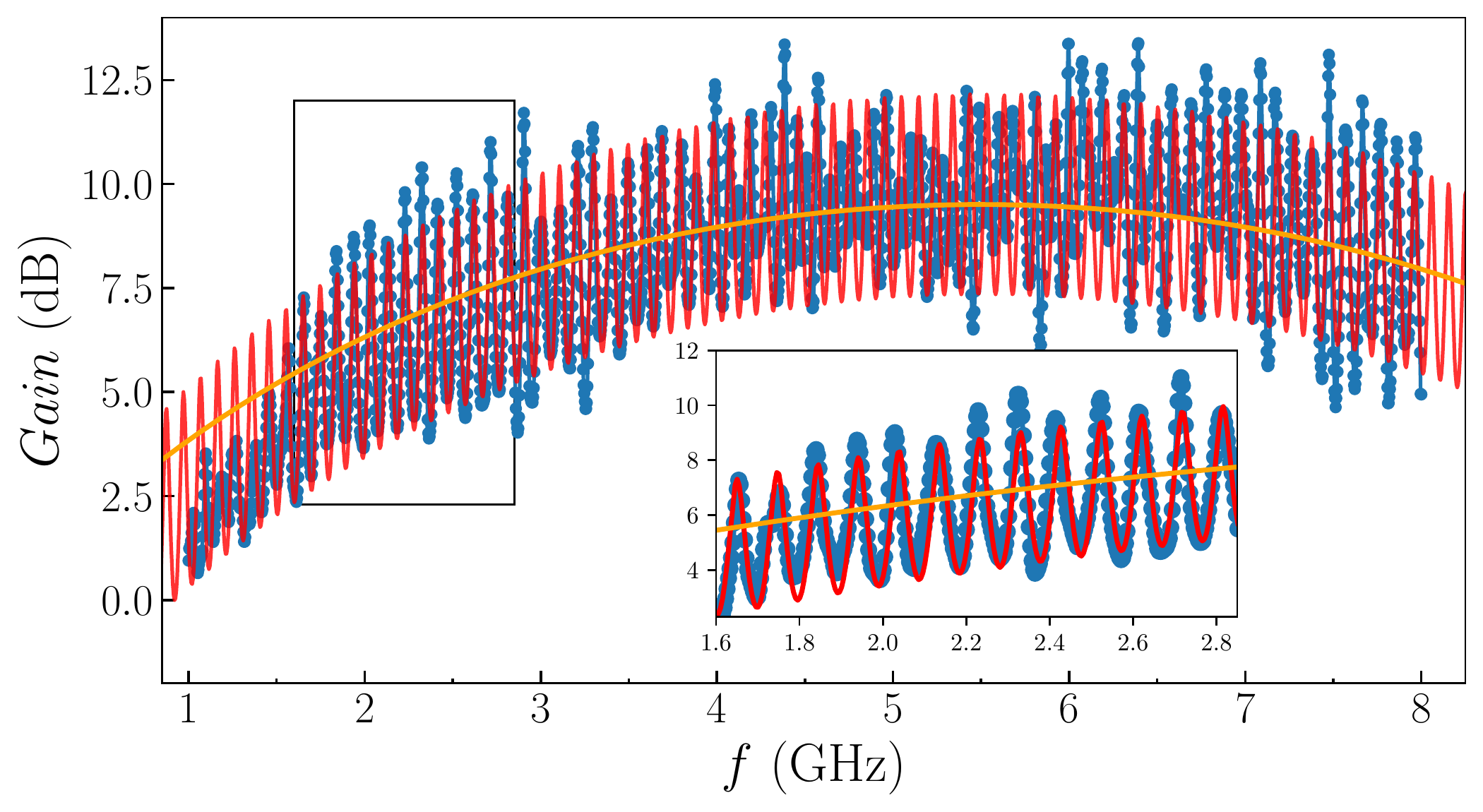}
	\caption{Measured amplification of kinetic inductance TWPA (blue data). Red curve is a fit by the modified theory. The
	inset shows how the Fabry-P\'{e}rot oscillations were matched to the measured transmission, allowing the extraction of phase
	velocity and reflection coefficient utilized to reconstruct the frequency profile of the gain. The orange curve is a fit of the data
	by standard theory.}
	\label{fig:sv}
\end{figure}

Fig.~\ref{fig:sv} presents the amplifier's gain measured in 3WM operation regime for DC bias current $I_D =1$~mA and the pump current amplitude $\mathcal{I}_p\approx1$~mA at frequency 10.38~GHz. These values were precisely chosen to pump the amplifier at the edge of the band gap at the highest possible pump power, which does not break superconductivity. At optimal values, the phase-matching condition $\beta_3=0$ is fulfilled and high gain is achieved. The measured gain profile shows an average gain of ~ 9.15 dB in the frequency range of 3-7~GHz, along with the presence of ripples indicating an impedance mismatch. As seen in Fig.~\ref{fig:sv} (inset), the ripples can be reasonably fitted with the proposed model. The reflection coefficient $\Gamma_p$ and phase velocity $v$  can be estimated from the fit, and are presented in Tab.~\ref{tab:params1} along with the values of $\mathcal{I}_p$ and $I_D$. 
For the kinetic inductance TWPA, $I_\ast$ is derived
within microscopic theory in Ref. \onlinecite{semenov2020effect} as $I_\ast\approx2I_c$. However, in experiments, deviations from this value were observed. For example, in Ref. \onlinecite{mahashabde2020fast}, for similar technology,
the ratio $I_\ast/I_c$ was estimated to be $\approx5$. This increase can be caused by the suppression of the critical current by vortex motion \cite{engel2008temperature}, and/or
by the presence of weak spots in the middle wire. In
other words, the average depairing current of the middle wire is higher than the critical current expected from the DC measurement. Therefore, here we express the parameters $\mathcal{I}_p$ and $I_D$
 in units of $I_\ast$. A fit to the standard CM theory returns slightly higher values of $I_D$ and $\mathcal{I}_p$ and lower
amplification.

\begin{table}[h]
	\begin{tabular}{l|>{\centering}p{1cm}|>{\centering}p{1cm}|>{\centering}p{1cm}|>{\centering}p{1cm}|p{1cm}|}
		\cline{2-6}
		& $l[\textrm{mm}]$	& $v[c]$                				 & $I_D[I_\ast]$                	 	& $\mathcal{I}_p[I_\ast]$	& \multicolumn{1}{>{\centering}p{1.5cm}|}{ $\Gamma_p$ } \\ \hline
		\multicolumn{1}{|l|}{standard theory} 	 		& \multirow{2}{*}{27} 	& \multirow{2}{*}{$0.018$} 			& 0.11					 	& 0.095  			&   \multicolumn{1}{>{\centering}p{1.5cm}|}{ 0} \\ \cline{1-1} \cline{4-6}
		\multicolumn{1}{|l|}{with correction} 			&                    		&                    					 & 0.1              		 		& 0.094			& \multicolumn{1}{>{\centering}p{1.5cm}|}{ 0.52}  \\ \hline
	\end{tabular}
	\caption{Parameters of the KI-TWPA. The length $l$ is given by the design, the other parameters are obtained from the fit of the amplification profile.}
	\label{tab:params1}
\end{table}

%%%%%%%%%%%%%%%%%%%%
\subsection{Josephson Junction TWPA }
%%%%%%%%%%%%%%%%%%%%
\begin{figure}
	\includegraphics[width=0.99\linewidth]{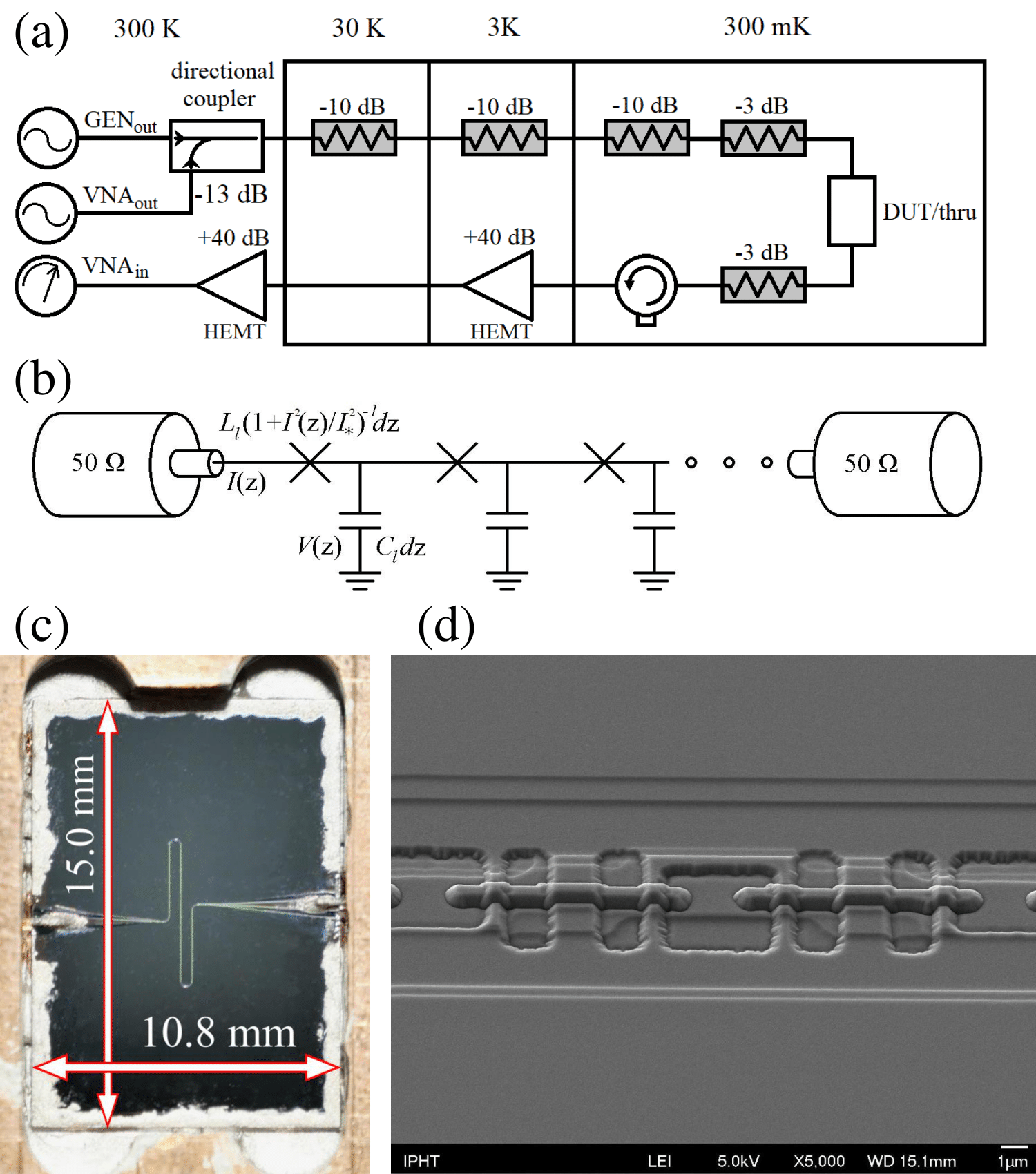}
	\caption{(a) Transmission measurement setup in refrigerator. The $3~$dB attenuator placed between
		the DUT and the isolator suppresses spurious resonances resulting from reflections between DUT and other parts of the setup. (b) Lumped element model of the JJ transmission line terminated by $50~\Omega$ coaxial
	cables. Estimated values of inductance $L_l$ and capacitance $C_l$ are listed in table
	\ref{tab:params2}. (c) Photo of the TWPA chip in copper box conected to SMA connector by indium. The sample is grounded by conductive silver varnish. (d) SEM image of Nb/AlO/Nb JJs forming the middle wire of TWPA.}
\label{fig:schemes}
\end{figure}
To show that the effect of gain enhancement of a TWPA occurs even for a higher reflection coefficient ($\Gamma_p\approx0.7$),
a high impedance CPW with nonlinear inductance was studied (see Fig. \ref{fig:schemes}). The middle wire, $11~$ mm long, is formed
out of 2000 niobium-based Josephson junctions (JJs),
see Appendix \ref{ap:device} as well as Ref.~\onlinecite{anders2009sub}. Assuming the inductance is dominated by the Josephson inductance, the inductance per unit length can be estimated by the relation $L_l l /2000 = \Phi_0/2\pi I_c$, where $\Phi_0$
is the magnetic flux quantum. Here, the critical current $I_c$ is estimated from the BCS relation for the product of $I_c$ and the normal state resistance of the junction $R_n$ \cite{ambegaokar1963tunneling}. The ground capacitance is estimated both by the standard formula (see Ref.~\onlinecite{goppl2008coplanar}), and an EM simulation in Sonnet software. These values, listed in Tab. \ref{tab:params2}, are utilized in the estimation of the phase velocity $1/\sqrt{L_lC_l}$ and the characteristic impedance $\sqrt{L_l/C_l}$. The design also contains photonic-crystal-like impedance modulation, however at the actual length of the waveguide, it has no observable effect.

The sample was installed in a copper box and its
$50~\Omega$ contacting pads were connected to the SMA connectors by indium (see (c) in Fig. \ref{fig:schemes}).
Prior to the actual measurement, the transmission has been calibrated by using a $50~\Omega$ TL connected instead of the TWPA (for
details on calibration, see Ref.~\onlinecite{kern2021transmission}) by a VNA in
a pulse-tube refrigerator at a temperature of $3.5~$K according to scheme shown in Fig. \ref{fig:schemes} (a).

\begin{figure}
	\includegraphics[width=\linewidth]{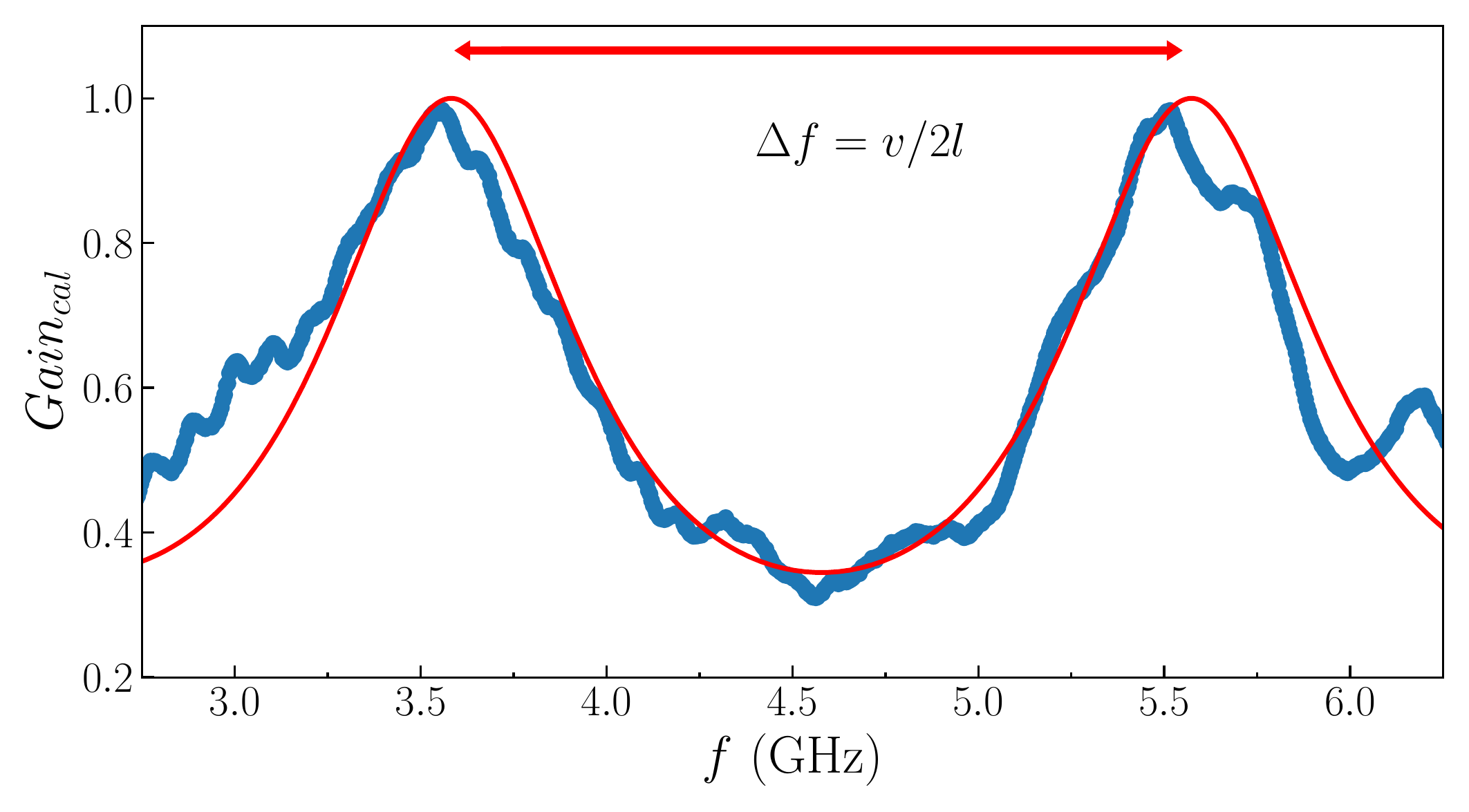}
	\caption{(a) Calibrated weak signal transmission without pumping tone (blue line). Fit of the data
	by Fabry-P\'{e}rot oscillating transmission (\ref{eq:Tran}) (red curve). The parameters of the fit; $\Gamma_{s}$ and $v$, are listed in table \ref{tab:params2}.}
	\label{fig:p_off_tran}
\end{figure}

Transmission measurement of the unmatched TL was performed at low signal power, where
nonlinearity is negligible (see blue line in Fig. \ref{fig:p_off_tran}).
The  measured transmission exhibits resonances, with peaks at frequencies $f_m=mv/2l$,
where $m$ is an integer, and $l$ is the length of the waveguide.
As the unmatched waveguide creates a stepped impedance resonator, the transmission can be described by Eq. (\ref{eq:Tran}) plotted as red line in Fig. \ref{fig:p_off_tran}.

\begin{figure}[h]
	\includegraphics[width=\linewidth]{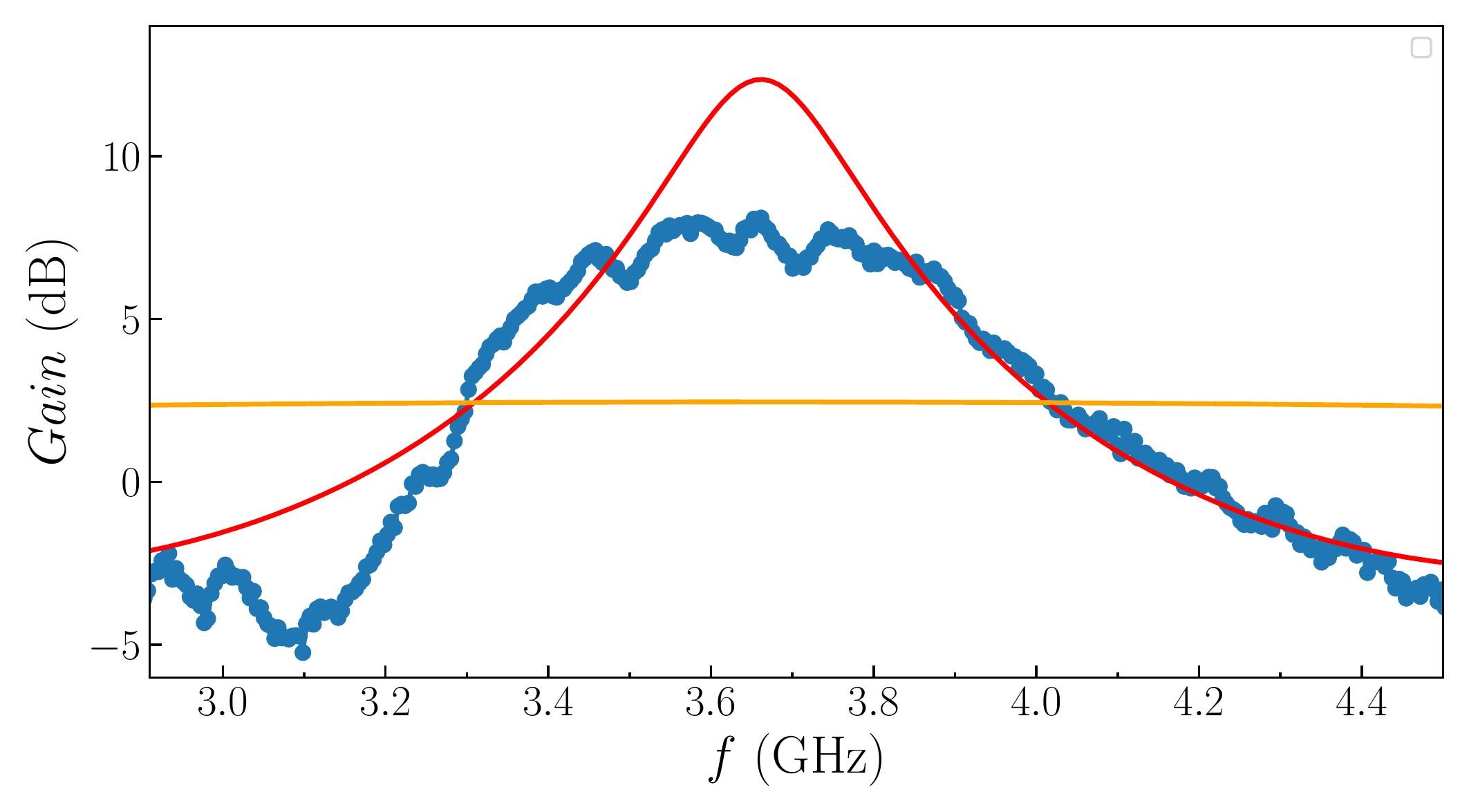}
	\caption{The measured gain of the TWPA (blue line) for pump frequency 3.38~GHz. Orange and red curves
	are results of the standard CME and the model with reflections for $\Gamma_p=0.72$, respectively. }
	\label{fig:cs}
\end{figure}

The obtained phase velocity $v$ and the reflection coefficient $\Gamma_{s}$ of the waveguide are consistent with the estimated inductance and capacitance per unit length (see Tab. \ref{tab:params2}). These parameters are used to calculate
the gain of the TWPA from Eq.~(\ref{eq:Ga}).
In addition, the reflection coefficient
$\Gamma_{s}$ is increased after the pump is applied and amplification is observed (see Appendix \ref{app:Q_inc}). 
Therefore, $\Gamma_{s}$ was replaced by
the value $\Gamma'_{s}\approx0.72$ obtained as a best fit to the experimental blue line by the theoretical red curve in Fig.~\ref{fig:cs}.

\begin{table}
	\centering
	
	 \begin{tabular}{|>{\centering}p{2cm}|>{\centering}p{2cm}>{\centering}p{2cm}>{\centering}p{2cm}}
		\hline
		$N$  								& \multicolumn{1}{c|}{$l~[\textrm{mm}]$}				 & \multicolumn{1}{c|}{$R_n~[\Omega]$} 		& \multicolumn{1}{c|}{$I_c~[\mu\textrm{A}]$} \\
		\hline
		2000         							& \multicolumn{1}{c|}{11}       							 & \multicolumn{1}{c|}{155}        						& \multicolumn{1}{c|}{12.3}\\
		\hline\hline
		$L_l~[\textrm{pH}/\mu\textrm{m}]$ 		& \multicolumn{1}{c|}{$C_l~[\textrm{fF}/\mu\textrm{m}]$} 	& \multicolumn{1}{c|}{$1/\sqrt{L_lC_l}[c]$} 				& \multicolumn{1}{c|}{$\sqrt{L_l/C_l}[\Omega]$}\\
		\hline
		4.68            							& \multicolumn{1}{c|}{0.13 }       						 & \multicolumn{1}{c|}{0.14}       						&\multicolumn{1}{c|} {189}\\
		\hline\hline
		$\Gamma_{s}$						& \multicolumn{1}{c|}{$\Gamma_{s}^\prime$}					 &\multicolumn{1}{c|} {$v~[c]$} 							& \multicolumn{1}{c|}{$\mathcal{I}_p[I_c]$} \\
		\hline
		0.51        							& \multicolumn{1}{c|}{0.72}       						 & \multicolumn{1}{c|}{0.14 }       						&\multicolumn{1}{c|} {0.84}\\
		\hline
	\end{tabular}
	\caption{Parameters of the JJ-TWPA waveguide. The length $l$ and the number of junctions $N$ are set by design. $NR_n$ is the room temperature resistance of the chain of $N$ JJs. Other parameters are obtained from the fit of the weak signal transmission by Eq. \ref{eq:Tran} and the amplified signal transmission by the presumed model.}
	\label{tab:params2}
\end{table}

The signal transmission with pump tone on was measured for various signal and pump powers. The highest amplification was achieved for pump power of -57.0~dBm and signal power ranging from -100~dBm to -86~dBm at the input of the device. Fig.~\ref{fig:cs} presents the gain as a function of the signal frequency, at the pump frequency 3.38 GHz. As there were no dispersion engineering features, and $\Delta k_4 = 0$ holds for all pump frequencies, a similar gain profile was obtained at various pump frequencies. The measured gain profile shows a region where the amplification occurs,
clearly corresponding to a peak resulting from resonance caused by the impedance mismatch (Fig. \ref{fig:p_off_tran}). When the standard
model for matched TL is applied ($\Gamma_p=0$, see orange curve in Fig.\ref{fig:cs}), a weak amplification ($<5~\textrm{dB}$) is
obtained over a wide bandwidth. Including the effects of reflections by the discussed model and accounting for the correction $\Gamma_p=\Gamma'_{s}$ from table
\ref{tab:params2} improves the correspondence between the theory and experiment. The ratio between the nonlinearity scale and the critical current for
JJs is more robust and it is given as $I_\ast/I_c=\sqrt{2}$. As the critical current estimated from the Josephson inductance agrees with the value calculated from the resistance of
junctions, the pump current amplitude is expressed in
units of $I_c$.

%------------------------------------------------
\section{Conclusion}
\label{ch:con}

In this article, we presented a modification of standard CME for parametric amplification,
which is commonly utilized to analyse TWPAs. By considering reflections due to impedance mismatches,
we showed that the gain ripples are an inherent property of unmatched finite length TWPAs.
Similarly to resonant amplifiers, in these peaks, the presented theory predicts enhanced gain. This feature can be utilized in the design  of shorter TWPAs. A shorter TL results in the broadening of gain ripples with bandwidth from few MHz up to 1 GHz and, at the same time, providing reasonable gain. The operating frequency range, where the unmatched TWPA provides gain in a series of Fabry-P\'{e}rot-like resonances is also increased by the reflections. Therefore, these peaks could be utilized in experiments requiring multiplexing.

Less demanding fabrication of the shorter TWPAs and omitting the challenging impedance matching may be advantages.
In this paper, therefore, we studied the working regime of  TWPA, in which the gain is increased by the reflections and the bandwidth is given by Fabry-P\'{e}rot resonances.  
We demonstrate the usability of our model by analyzing the response of two types of devices: a JJ and a kinetic inductance TWPA. For the
kinetic inductance TWPA $27~$mm long with slight impedance mismatch, good agreement with the experiment  was
achieved. Series of $30-40~$MHz wide peaks with gain over $10~$dB are observed in range from $3-7~$GHz.
The JJ TPWA was significantly shorter $(l=11~$mm$)$ with higher reflections providing $8~$dB gain in a single peak with bandwidth of $\approx600~$MHz.

\section{Acknowledgments}
We thank L. Planat and N. Roch (Institut N\'{e}el, CNRS, Grenoble) and D. M. Basko (
Universit\'{e} Grenoble Alpes, CNRS, LPMMC, Grenoble) for useful discussions.
This work was supported by the European Unions Horizon 2020 research and innovation
programme under Grant Agreement No. 863313 (SUPERGALAX), No. 362660 (Quantum E-Leaps), and by the SAS-MVTS, Grant QuantERA-SiUCs..
The support from the Slovak Research and Development Agency under the contracts APVV-16-0372, and
APVV-20-0425 are gratefully acknowledged.
The Chalmers group acknowledges the support from the Swedish Research Council (VR) (grant agreements 2016-04828 and 2019-05480), EU H2020 European Microkelvin Platform (grant agreement 824109), Engineering and Physical Sciences Research Council (EPSRC) Grant No. EP/T004088/1 and from Knut and Alice Wallenberg Foundation via the Wallenberg center for Quantum Technology (WACQT).
This work has relieved funding from German Federal Ministry of Education and Research (BMBF) under the project QSolid (Grant No. 13N16152) and the Free State of Thuringia under No. 2021 FGI 0049.
\appendix
%%%%%%%%%%%%%%%%%%%%
\section{Coupled mode equations for waves in a resonator}
\label{ap:CM}
%%%%%%%%%%%%%%%%%%%%
The field inside a Fabry-P\'{e}rot resonator terminated by two identical
mirrors with reflection coefficients $\Gamma_n$ consists of an infinite number of reflected waves.
Thus the amplitude $I_n$ of the field oscillating at frequency $\omega_n$ is defined as
\begin{equation}
\label{eq:reflections}
\begin{aligned}
I_n(z,t)=&\frac{1}{2}\mathcal{I}_n(z)(e^{ik_nz}+\Gamma_n e^{ik_n(l-z)}+\\
+\Gamma_n^2&e^{ik_n(2l+z)}+...)e^{-i\omega_n t}+c.c=\\
&=\frac{1}{2}\mathcal{I}_n(z)t_n(e^{ik_nz}+\tilde{\Gamma_n} e^{-ik_nz})e^{-i\omega_n t}+c.c.,
\end{aligned}
\end{equation}
where in the last line, the following notation is introduced

\noindent\begin{minipage}{.49\linewidth}
\begin{equation}
\label{eq:gamma_n}
\tilde{\Gamma}_n = \Gamma_n e^{ik_nl},
\end{equation}
\end{minipage}
\begin{minipage}{.5\linewidth}
\begin{equation}
\label{eq:t_n}
t_n=\frac{1}{1-\tilde{\Gamma}_n^2}.
\end{equation}
\end{minipage}
Solving the nonlinear wave equation \eqref{eq:WE} for current in the form $I(z,t)=I_s(z,t)+I_p(z,t)+I_{i_3}(z,t)+I_{i_4}(z,t)$ means finding the spatially-dependent current amplitudes
$\mathcal{I}_n(z)$, such that the current \eqref{eq:waves2}, satisfies
\eqref{eq:WE}. Substituting \eqref{eq:waves2} into
\eqref{eq:WE} and adopting notation $\mathcal{I}'_n\equiv\frac{\partial \mathcal{I}_n}{\partial z}$ one obtains
within a slowly varying envelope approximation (i.e. $\left|\mathcal{I}''_n\right|\ll k_n\left|\mathcal{I}'_n\right|$) the following equation
\begin{equation}
\label{eq:the_eq}
\begin{aligned}
\frac{1}{L_lC_l}ik_n\mathcal{I}'_n(z)t_n(e^{ik_nz}-\tilde{\Gamma}_n e^{-ik_nz})=&\\
-\frac{\epsilon\omega_n^2}{2}\sum_{\substack{a,b\in \\ \{p,s,i_3\}}} \Big(\prod_{\substack{m \in \\ \{a,b\}}}\frac{1}{2}\mathcal{I}_m^{\pm}t_m^{\pm}(e^{\pm ik_mz}+\tilde{\Gamma}_m^\pm& e^{\mp ik_mz})\Big)\\\times\delta(\omega_n\mp\omega_a&\mp\omega_b)\\
-\frac{\xi\omega_n^2}{3}\sum_{\substack{a,b,c\in \\ \{p,s,i_4\}}} \Big(\prod_{\substack{m \in \\ \{a,b,c\}}}\frac{1}{2}\mathcal{I}_m^{\pm}t_m^{\pm}(e^{\pm ik_mz}+\tilde{\Gamma}_m^\pm& e^{\mp ik_mz})\Big)\\\times\delta(\omega_n\mp\omega_a\mp\omega_b&\mp\omega_c),\\
\end{aligned}
\end{equation}
where the first sum on the right hand side describes 3WM terms and the second includes 4WM terms.
Here, $\mathcal{I}_m^+=\mathcal{I}_m$, $t_m^+=t_m$ and $\tilde{\Gamma}_m^+=\tilde{\Gamma}_m$ whereas $\mathcal{I}_m^-=\mathcal{I}_m^\ast$, $t_m^-=t_m^\ast$ and $\tilde{\Gamma}_m^-=\tilde{\Gamma}_m^\ast$ i.e. minus in superscript indicates complex conjugation. 

To study the parametric amplification provided by 3 and 4-wave mixing the equation (\ref{eq:the_eq})
is solved for four waves: strong pump $\mathcal{I}_p$, signal $\mathcal{I}_s$ and two idlers
$\mathcal{I}_{i_{3,4}}$ such that $\mathcal{I}_p\gg \mathcal{I}_s, \mathcal{I}_{i_{3,4}}$. Making use of conditions (\ref{eq:EC3}) and (\ref{eq:EC4}) in the delta functions in equation (\ref{eq:the_eq}), one obtains
\begin{equation}
\label{eq:pump}
\mathcal{I}'_p=\frac{ik_p}{8}\xi\left|\mathcal{I}_p\right|^2\left|t_p\right|^2\mathcal{I}_p\mathcal{F}^{pp}_{pp},
\end{equation}
\begin{equation}
\begin{aligned}
\label{eq:signal}
\mathcal{I}'_s=\frac{ik_s}{4}&\epsilon\mathcal{I}_p\mathcal{I}_{i_3}^{\ast}\frac{t_pt_{i_3}^\ast}{t_s}\mathcal{F}^{p}_{i_3s}\\
+\frac{ik_s}{8}\xi\Big(\mathcal{I}_p^2\mathcal{I}_{i_4}^{\ast}\frac{t_p^2t_{i_4}^\ast}{t_s}&\mathcal{F}^{pp}_{i_4s}+
2\left|\mathcal{I}_p\right|^2\mathcal{I}_s\left|t_p\right|^2\mathcal{F}^{ps}_{ps}\Big),
\end{aligned}
\end{equation}
\begin{equation}
\label{eq:idler3}
\mathcal{I}'_{i_3}=\frac{ik_{i_3}}{4}\Big(\epsilon\mathcal{I}_p\mathcal{I}_s^{\ast}\frac{t_pt_s^\ast}{t_{i_3}}\mathcal{F}^{p}_{si_3}+
\xi\left|\mathcal{I}_p\right|^2\mathcal{I}_{i_3}\left|t_p\right|^2\mathcal{F}^{pi_3}_{pi_3}\Big),
\end{equation}
\begin{equation}
\label{eq:idler4}
\mathcal{I}'_{i_4}=\frac{ik_{i_4}}{8}\xi\Big(\mathcal{I}_p^2\mathcal{I}_s^{\ast}\frac{t_p^2t_s^\ast}{t_{i_4}}\mathcal{F}^{pp}_{si_4}+
2\left|\mathcal{I}_p\right|^2\mathcal{I}_{i_4}\left|t_p\right|^2\mathcal{F}^{pi_4}_{pi_4}\Big),
\end{equation}
where
\begin{equation}
\label{eq:defF}
\begin{aligned}
\mathcal{F}^{ab}_{cd}(z) = (e^{ik_az}+\tilde{\Gamma}_a e^{-ik_a z})(e^{ik_bz}+\tilde{\Gamma}_b e^{-ik_b z})\\\times\frac{(e^{-ik_c z}+{\tilde{\Gamma}_c}^\ast e^{ik_c z })}{e^{ik_dz}-\tilde{\Gamma}_d e^{-ik_dz}}\quad a,b,c,d\in \{p,s,i_3,i_4\},
\end{aligned}
\end{equation}
\begin{equation}
\label{eq:defF}
\begin{aligned}
\mathcal{F}^{a}_{cd}(z) = (e^{ik_az}+\tilde{\Gamma}_a e^{-ik_a z})\frac{(e^{-ik_b z}+{\tilde{\Gamma}_b}^\ast e^{ik_b z })}{e^{ik_cz}-\tilde{\Gamma}_c e^{-ik_cz}}\\ a,b,c\in\{p,s,i_3\}.
\end{aligned}
\end{equation}
The equation for pump (\ref{eq:pump}) is solved in depleted pump approximation $\left|\mathcal{I}_p\right|'=0$:
\begin{equation}
\label{eq:pumpres}
\mathcal{I}_p=\left|\mathcal{I}_p\right|\textrm{exp}(i\int_0^z\overset{\approx}\kappa_{p}\mathcal{F}^{pp}_{pp}dx),
\end{equation}
where
\begin{equation}
\label{eq:kappa_n}
\overset{\approx}\kappa_{n}=k_n\frac{\left|\mathcal{I}_p\right|^2t_p^2}{8}\xi=k_n\frac{\left|\mathcal{I}_p\right|^2t_p^2}{8}\frac{1}{(I_{D}^2+I_\ast^2)},
\end{equation}
\begin{equation}
\label{eq:kappa_nD}
\overset{\simeq}{\kappa}_{n}=k_n\frac{t_p\mathcal{I}_p}{4}\epsilon=k_n\frac{t_p\mathcal{I}_p}{4}\frac{2I_{D}}{(I_{D}^2+I_\ast^2)}.
\end{equation}
Although the exponential in eq. (\ref{eq:pumpres}) contains also a real part, changing the module
$\left|\mathcal{I}_p\right|$ too, the contribution oscillates at the scale of $1/k_p$ and it is smaller than imaginary part by factor $\Gamma_p$, thus it is neglected.
Utilizing solution (\ref{eq:pumpres}), the equations (\ref{eq:signal}), (\ref{eq:idler3}) and (\ref{eq:idler4})
are simplified by the following transformation:
\begin{equation}
\label{eq:itoa}
A_n = \mathcal{I}_nt_n\textrm{exp}\big(-2i\int\displaylimits_0^z\overset{\approx}\kappa_n\mathcal{F}^{pn}_{pn}dx\big)\quad n=s,i_3,i_4.
\end{equation}
Finally, one obtains the coupled mode equations for the complex amplitudes $A_s$  and
$A_{i_{3,4}}$:
\begin{equation}
\label{eq:das}
\begin{aligned}
A'_s=i\overset{\simeq}{\kappa}_sA_{i_3}^\ast\mathcal{F}^{p}_{i_3s}e^{ib_3}\\+i\overset{\approx}\kappa_sA_{i_4}^\ast\mathcal{F}^{pp}_{i_4s}e^{ib_4},
\end{aligned}
\end{equation}
\begin{equation}
\label{eq:dai3}
A'_{i_3}=i\overset{\simeq}{\kappa}_{i_3}A_s^\ast\mathcal{F}^{p}_{si_3}e^{ib_3},
\end{equation}
\begin{equation}
\label{eq:dai4}
A'_{i_4}=i\overset{\approx}\kappa_{i_4}A_s^\ast\mathcal{F}^{pp}_{si_4}e^{ib_4}.
\end{equation}
Here, the functions $b_3(z)$ and $b_4(z)$ contain contributions to the nonlinear phase modulations and are expressed as follows
\begin{equation}
\label{eq:bz}
b_3(z)=\int_0^z \Big(\overset{\approx}\kappa_p\mathcal{F}^{pp}_{pp}(x)-2\overset{\approx}\kappa_s \mathcal{F}^{ps}_{ps}(x) -2\overset{\approx}\kappa_{i_3} \mathcal{F}^{pi_3}_{pi_3}(x)\Big)dx,
\end{equation}
\begin{equation}
\label{eq:bz}
b_4(z)=2\int_0^z \Big(\overset{\approx}\kappa_p\mathcal{F}^{pp}_{pp}(x)- \overset{\approx}\kappa_s \mathcal{F}^{ps}_{ps}(x) - \overset{\approx}\kappa_{i_4} \mathcal{F}^{pi_4}_{pi_4}(x)\Big)dx.
\end{equation}
In the following subsections, the obtained equations are solved in two cases: 1. $I_D\gg\mathcal{I}_p$ (3-wave mixing) \cite{malnou2021three,erickson2017theory} and 2. $I_D=0$ (4-wave mixing).

\subsection{3-wave mixing}
To study 3WM, let us assume that the DC bias current is much larger than the pump amplitude, which means, according to eq. (\ref{eq:kappa_n}, \ref{eq:kappa_nD}), that $\overset{\simeq}{\kappa}_n\gg\overset{\approx}\kappa_n$ .
If no idler is applied at the frequency $\omega_{i_4}$ to the input of the device, the 4WM idler is generated proportionally to $\overset{\approx}\kappa_{i_4}$. Therefore, the 4WM idler is much weaker than 3WM idler which is proportional to $\overset{\simeq}{\kappa}_{i_3}$ and the system of equations (\ref{eq:das}-\ref{eq:dai4}) is approximated by two coupled equations
\begin{equation}
\label{eq:das33}
\begin{aligned}
A'_s=i\overset{\simeq}{\kappa}_sA_{i_3}^\ast\mathcal{F}^{p}_{i_3s}e^{ib_3},
\end{aligned}
\end{equation}
\begin{equation}
\label{eq:dai33}
A'_{i_3}=i\overset{\simeq}{\kappa}_{i_3}A_s^\ast\mathcal{F}^{p}_{si_3}e^{ib_3},
\end{equation}
which can be easily decoupled. For $A_s$ the uncoupled equation is
\begin{equation}
\label{eq:3evolution}
A_s'' - \frac{ (e^{ib_3}\mathcal{F}^{p}_{i_3s})'}{e^{ib_3}\mathcal{F}^{p}_{i_3s}}A_s' - \overset{\simeq}{\kappa}_s\mathcal{F}^{p}_{i_3s}\Big({\overset{\simeq}{\kappa}_{i_3}}{\mathcal{F}^{p}_{si_3}}\Big)^\ast A_s e^{-i2\mathfrak{Im}(b_3)}= 0.
\end{equation}

\subsection{4-wave mixing}
When no DC bias is applied, pure 4WM is observed and the system (\ref{eq:das} - \ref{eq:dai4}) becomes
\begin{equation}
\label{eq:das44}
\begin{aligned}
A'_s=i\overset{\approx}\kappa_sA_{i_4}^\ast\mathcal{F}^{pp}_{i_4s}e^{ib_4},
\end{aligned}
\end{equation}
\begin{equation}
\label{eq:dai44}
A'_{i_4}=i\overset{\approx}\kappa_{i_4}A_s^\ast\mathcal{F}^{pp}_{si_4}e^{ib_4},
\end{equation}
which gives the equation for the spatial evolution of the transformed signal amplitude $A_s(z)$:
\begin{equation}
\label{eq:4evolution}
A_s'' - \frac{ (e^{ib_4}\mathcal{F}^{pp}_{i_4s})'}{e^{ib_4}\mathcal{F}^{pp}_{i_4s}}A_s' - \overset{\approx}\kappa_s\mathcal{F}^{pp}_{i_4s} \Big(\overset{\approx}\kappa_{i_4}{\mathcal{F}^{pp}_{si_4}}\Big)^\ast A_s e^{-i2\mathfrak{Im}(b_4)}= 0,
\end{equation}

A simple check of our derivation is achieved by identifying equations (\ref{eq:3evolution},
\ref{eq:4evolution}) as the general form of the standardly presented CME result for $\Gamma_n=0$.
%%%%%%%%%%%%%%%%%%%%
\section{Approximation of the gain equation}
\label{ch:app}
%%%%%%%%%%%%%%%%%%%%
Equations (\ref{eq:3evolution}, \ref{eq:4evolution}) for slow variation of an envelope
of waves propagating and reflecting in a nonlinear Fabry-P\'{e}rot resonator are second-order
differential equations where  $\mathcal{F}$ and $b$ are functions of $z$. To solve the
differential equations, we expand these functions up to second order in the reflection
coefficient $\Gamma_n<1$ and remove terms with harmonic spatial dependence via the
averaging method \cite{sanders2007averaging}. This transformation is indicated by the
arrows in the equations given below. This procedure yields an equation similar to the result
of the standard CME for waves propagating only in one direction:
\begin{equation}
\label{eq:Fpppp}
\begin{aligned}
\mathcal{F}^{pp}_{pp}\approx& 1+2\Gamma_pe^{-ik_p(2z-l)}+2\mathfrak{Re}(\Gamma_p e^{-ik_p(2z-l)})\\&+3\left|\Gamma_p\right|^2  +4\Gamma_p^2 e^{-i2k_p(2z-l)}\rightarrow 1+3\left|\Gamma_p\right|^2
\end{aligned}
\end{equation}
\begin{equation}
\label{eq:Fpnpn}
\begin{aligned}
\mathcal{F}^{pn}_{pn}\approx& 1+2\Gamma_n e^{-ik_n(2z-l)}+2\Gamma_n^2e^{-ik_s(2z-l)}\\
&+2\mathfrak{Re}(\Gamma_p e^{-ik_p(2z-l)})(1+2\Gamma_n e^{-i(k_n)(2z-l)})\\
&+ \left|\Gamma_p\right|^2\rightarrow 1+\left|\Gamma_p\right|^2\quad n=s,i_4,
\end{aligned}
\end{equation}
\begin{equation}
\label{eq:FDpsi3}
\begin{aligned}
{\mathcal{F}^{p}_{si_3}}^\ast\mathcal{F}^{p}_{i_3s}\rightarrow 1+\left|\Gamma_p\right|^2,
\end{aligned}
\end{equation}
\begin{equation}
\label{eq:Fppsi4}
\begin{aligned}
{\mathcal{F}^{pp}_{si_4}}^\ast\mathcal{F}^{pp}_{i_4s}\rightarrow 1+4\left|\Gamma_p\right|^2.
\end{aligned}
\end{equation}

The above approximations leads to the equation describing 3WM
\begin{equation}
\label{eq:fin}
A_s'' - i\beta_3 A_s' - \overset{\simeq}{\kappa}_s{\overset{\simeq}{\kappa}_{i_3}^\ast}(1+\left|\Gamma_p\right|^2) A_s = 0
\end{equation}
which gives the nonlinear phase mismatch $\beta_3$
\begin{equation}
\label{eq:beta_gamma_p}
\beta_3=\Delta k_3+\overset{\approx}\kappa_p( 1+3\left|\Gamma_p\right|^2)-2\overset{\approx}\kappa_s( 1+\left|\Gamma_p\right|^2)-2\overset{\approx}\kappa_{i_3}( 1+\left|\Gamma_p\right|^2)),
\end{equation}
where $\Delta k_3=k_p-k_s-k_{i_3}$ is the deviation from linear dispersion relation.

Similarly, the differential equation for 4WM takes the form
\begin{equation}
\label{eq:fin}
A_s'' - i\beta_4 A_s' - \overset{\approx}\kappa_s\overset{\approx}\kappa_{i_4}^\ast(1+4\left|\Gamma_p\right|^2) A_s = 0
\end{equation}
and the nonlinear phase mismatch $\beta_4$ reads
\begin{equation}
\label{eq:beta_gamma_p}
\beta_4=\Delta k_4+2(\overset{\approx}\kappa_p( 1+3\left|\Gamma_p\right|^2)-\overset{\approx}\kappa_s( 1+\left|\Gamma_p\right|^2)-\overset{\approx}\kappa_{i_4}( 1+\left|\Gamma_p\right|^2)),
\end{equation}
where $\Delta k_4=2k_p-k_s-k_{i_{4}}$.

With the boundary conditions $\mathcal{I}_s(z=0)=\mathcal{I}_{s0}$ and $\mathcal{I}_{i_{3,4}}(z=0)=0$ the solution takes the compact form
\begin{equation}
\begin{aligned}
\label{eq:is_vs_z}
\mathcal{I}_s(z)=\mathcal{I}_{s0}\Big(\textrm{cosh}(g_{3,4}z)-i\frac{\beta_{3,4}}{2g_{3,4}}\textrm{sinh}(g_{3,4}z)\Big)\\
\times e^{i(\frac{\beta_{3,4}}{2}+2\overset{\approx}\kappa_s(1+\left|\Gamma_p\right|^2))z},
\end{aligned}
\end{equation}
where
\begin{equation}
\label{eq:is_vs_z}
g_3=\sqrt{\overset{\simeq}{\kappa}_s{\overset{\simeq}{\kappa}_{i_3}^\ast}(1+\left|\Gamma_p\right|^2)-\frac{\beta_3^2}{4}},
\end{equation}
\begin{equation}
\label{eq:is_vs_z}
g_4=\sqrt{\overset{\approx}\kappa_s\overset{\approx}\kappa_{i_4}^\ast(1+4\left|\Gamma_p\right|^2)-\frac{\beta_4^2}{4}}
\end{equation}
are gain factors for 3WM and 4WM, respectively.
%%%%%%%%%%%%%%%%%%%%
\section{Quality increase due to amplification}
\label{app:Q_inc}
%%%%%%%%%%%%%%%%%%%%
Following the conventional derivation of the reflection coefficient $\Gamma_n$ at the ends of the
TL (see Ref.~\onlinecite{pozar2011microwave}), we derive below the influence of the amplification
on the reflection coefficient and, therefore, on the quality factor of the resonator. Utilizing
the first telegrapher equation (\ref{eq:TEV}) for harmonic components:
\begin{equation}
\label{eq:h_TE_1}
I_n'(z)=-i\omega_sCV_n(z),
\end{equation}
the voltage amplitude along the waveguide is found by substituting the current amplitude, which
was determined in the previous chapter (\ref{ap:CM}).
This way the impedance at the end of the waveguide is obtained:
\begin{equation}
\label{eq:Zl}
Z_n(l)=\frac{V_s(l)}{I_s(l)}\approx Z_0\Big(1+i\frac{\mathcal{I}_n(z)'|_{z=l}}{\mathcal{I}_n(l)k_n}\Big),
\end{equation}
%\vspace{0.1cm}
where $Z_0=\sqrt{L_l/C_l}$ and the nonlinear spatial variation of phase velocity was neglected.
Finally, the reflection coefficient can by expressed as
\begin{equation}
\label{eq:gamma_vs_amp}
\Gamma_n=\frac{Z_L-Z_0-Z_0i\frac{\mathcal{I}_n(z)'|_{z=l}}{\mathcal{I}_n(l)k_n}}{Z_L+Z_0+Z_0i\frac{\mathcal{I}_n(z)'|_{z=l}}{\mathcal{I}_n(l)k_n}},
\end{equation}
showing that the reflection is sensitive to any spatial variation in the amplitude
of the current.

\section{Samples preparation}
\label{ap:device}
The 140 nm NbN film for a KI-TWPA was fabricated at Chalmers, following the recipe Ref.~\onlinecite{mahashabde2020fast}.
These films were deposited on a sapphire wafer and the desired structure was patterned with
e-beam lithography and Ar:Cl$_2$ plasma etching. The central line was further thinned down to 30~nm,
to get a sheet resistance of $\sim$ 48~$\Omega$/$\square$, which corresponds to the desired kinetic
inductance of $\sim$ 4~pH/$\square$. The thickness of the ground plane and other elements was left
unchanged, in order to keep the KI of these elements low and thereby eliminate self-resonances in
fractal structures.

The JJ devices were fabricated at Leibniz IPHT by making use of the so-called cross-type Josephson junction technology.
Here a trilayer of Nb/AlO$_x$/Nb with a critical current density of about 1.7~kA/cm$^2$ is deposited
on an oxidized 4 inch silicon wafer of 500~$\mu$m thickness. Thermal oxide thickness on the wafer
was about 600~nm. Inside a meander shaped Nb groundplane with a 9~$\mu$m slit, an array of Josephson
junctions form the center conductor. In total, 2000 Josephson junctions with a nominal junction size
of ($0.9\times0.9$)~$\mu$m$^{2}$ are fabricated on a single chip, with dimensions of ($10800\times15000$)
 ~$\mu$m$^{2}$. By means of Fiske step measurements, the specific junction capacitance has been determined
to be around 60~fF/$\mu$m$^{2}$ for this critical current density. For sample fabrication, electron beam
lithography has been used. Nb patterning was done by making use of reactive ion etching based on CF$_{4}$.

%----------------------------------------------------------------------------------------
%	REFERENCE LIST
%----------------------------------------------------------------------------------------

%----------------------------------------------------------------------------------------
\end{document}